\documentclass{elsart5p}

\usepackage{natbib}
\usepackage{xspace}

\usepackage{epsfig}

\usepackage{amssymb}

\def\cref#1{Chapt.\,\ref{#1}}
\def\Cref#1{Chapter~\ref{#1}}

\def\fref#1{Fig.\,\ref{#1}}
\def\ffref#1{Figs.\,\ref{#1}}

\def\tref#1{Table~\ref{#1}}

\newlength{\xx}
\begin{document}

\begin{frontmatter}

\title{Forbush decreases and solar events seen in the $10-20$~GeV energy range\\
 by the Karlsruhe Muon Telescope}

\author{I. Braun$^{a,1}$, J. Engler$^b$, J.R. H\"orandel$^{a,2}$, and J.
Milke$^{b,3}$}

\address{$^a$Institut f\"ur Experimentelle Kernphysik, Universi\"at
Karlsruhe, D-76021 Karlsruhe, Germany\\
$^b$ Institut f\"ur Kernphysik, Forschungszentrum Karlsruhe, D-76021
Karlsruhe, Germany}

\thanks{corresponding author, email:Isabel.Braun@phys.ethz.ch,
now at: Institute for Particle Physics, Schafmattstr. 20, 8093 Z\"urich, Switzerland}

\thanks{now at: Radboud University Nijmegen, Department of
Astrophysics, P.O. Box 9010, 6500 GL Nijmegen, The Netherlands}

\thanks{now at: Institut f\"ur Wissenschaftliches Rechnen,
Forschungszentrum Karlsruhe, D-76021 Karlsruhe, Germany}

\journal{Advances in Space Research}

\begin{abstract} 
 Since 1993, a muon telescope located at Forschungszentrum Karlsruhe (Karlsruhe
 Muon Telescope) has been recording the flux of single muons mostly originating
 from primary cosmic-ray protons with dominant energies in the $10-20$~GeV
 range.  The data are used to investigate the influence of solar effects on the
 flux of cosmic-rays measured at Earth.  Non-periodic events like Forbush
 decreases and Ground Level Enhancements are detected in the registered muon
 flux.  A selection of recent events will be presented and compared to data
 from the Jungfraujoch neutron monitor. The data of the Karlsruhe Muon
 Telescope help to extend the knowledge about Forbush decreases and Ground
 Level Enhancements to energies beyond the neutron monitor regime.
\end{abstract}


\end{frontmatter}

\parindent=0.5 cm

\section{Introduction}

The association of solar activity with the cosmic-ray intensity has been
studied for various observed effects including Forbush decreases \citep{For54},
i.e.\ a rapid decrease in the observed galactic cosmic-ray intensity, and
Ground Level Enhancements, which are connected to large solar flares. They can
be related to magnetic disturbances in the heliosphere that create transient
cosmic-ray intensity variations \citep{Par65,kallenrode}. From the observation
of such events with different experiments, an energy dependent description can
be obtained.  The heliospheric influence is mostly pronounced for primary
particles with low rigidity and has been studied mainly using data of the
worldwide neutron monitor network \citep{Sim00}.  With its unique median
primary energy of 40~GeV for protons, the Karlsruhe Muon Telescope fills the
energy gap between neutron monitors (from $\approx 11-15$~GeV, depending on solar 
activity state, to $\approx33$~GeV) and other muon
telescopes ($\approx 53-119$~GV rigidity). In the following, we report on the
detection of Forbush decreases and the investigation of Ground Level
Enhancements with the Karlsruhe Muon Telescope.

\section{Experimental Set-Up}

\begin{figure}\centering
 \epsfig{width=\columnwidth,file=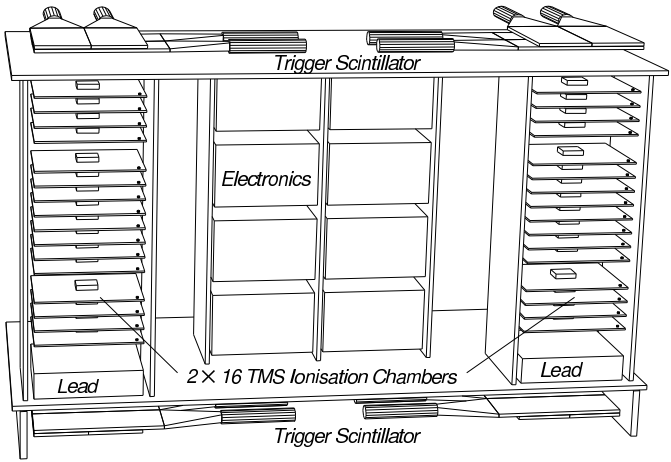}
 \caption{Sketch of the muon telescope.  It consists of two "towers", each
	  comprising two double layers of scintillation counters (arranged in a
	  $2\times2$ matrix), separated by a 16~cm lead absorber.}
 \label{fig0}
\end{figure}

The flux of single muons from the zenith region has been recorded continuously
since 1993 with the Karlsruhe Muon Telescope located at Forschungszentrum
Karlsruhe, Germany (49.094$^\circ$N, 8.431$^\circ$E, 120~m a.s.l). The set-up
is sketched in \fref{fig0}, details are given by \citet{Eng99}.  Two double
layers of scintillation counters are arranged on top of each other, separated
by a 16~cm lead absorber, forming a "tower".  Each scintillation counter
comprises a scintillator (NE~102) with the dimensions $0.6~\mbox{m} \times
0.25~\mbox{m}\times 0.02$~m, read out by a photomultiplier via an adiabatic
light guide.  A double layer is formed by two scintillation counters arranged
parallel to each other and two counters oriented perpendicular to them, forming
a $2\times2$ detector matrix.  The lead absorber selects muons with energies
larger than 0.8~GeV.  Two such towers with a separation of 1.8~m are operated
with a veto trigger logic selecting vertical particles and rejecting showers in
which more than one particle hits the detector, thus suppressing the hadronic
background  to about 0.8\% of the events \citep{kfk5320}. The muon detector is
operated in a climatized room at a stable temperature.  The towers of the
instrument can be used to calibrate up to 32 liquid ionization chambers for the
KASCADE hadron calorimeter, their data are not included in the present
analysis.  This analysis includes 80\,017~h of data between October 1993 and
November 2006.

\begin{figure}\centering
 \epsfig{width=\columnwidth,file=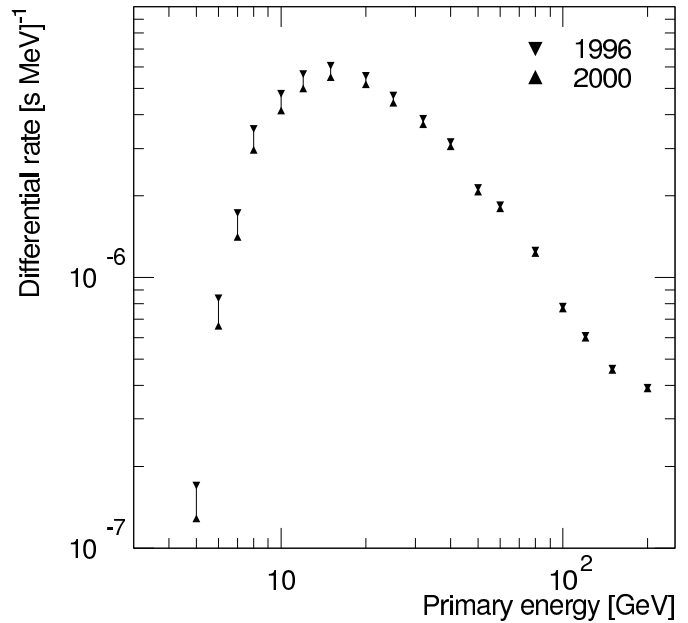}
 \caption{Simulated differential trigger rate of the Karlsruhe Muon Telescope
	  as a function of primary energy for solar minimum (1996) and maximum
          (2000).}
 \label{fig1}
\end{figure}

From simulations with CORSIKA \citep{corsika} and GEANT 3.21 \citep{geant},
properties of the primary particles were investigated.  The simulations include
the physics processes in the atmosphere and the propagation through the
building surrounding the detector, as well as the trigger conditions of the
muon telescope. It turns out that the muons originate mostly from cosmic-ray
protons, of which 95\% have zenith angles smaller than $18^\circ$. The
differential energy spectrum of primary protons triggering the telescope was
derived for parameterized primary spectra in different solar activity states
\citep{Urc72}, taking into account modulation parameters according to
\citet{Uso03}.  The differential trigger rate obtained, in other words the 
detector response function folded with the
primary particle spectrum, is presented in
\fref{fig1}.  The maximum occurs at primary energies of about 15~GeV.  The
expected counting rate difference between the two activity states is 2.5\%.
The median energy $E_M$ of a detector is defined such that one half of the
detected events originate from primary particles below (or above) $E_M$. The
median energy of the Karlsruhe Muon Telescope for both simulated spectra is
40~GeV.  

\section{Atmospheric Corrections}

Muons loose energy and decay on the way from their production site in the
atmosphere to the detector, yielding a dependence of the detected rate on the
height of the production layer and the amount of material traversed above the
detector. Corrections were applied to the recorded muon rate using the
atmospheric pressure measured at the Forschungszentrum Karlsruhe.  In addition,
balloon ascends at noon and midnight conducted by the German weather service
(DWD) in Stuttgart provide the heights of specific pressure layers including
the 150~g/cm$^2$ layer ($\approx13.6$~km) which is close to the typical
production layer of muons triggering the telescope at 130~g/cm$^2$, as
determined from simulations.

For each year, the muon rate was iteratively corrected for a pressure of
1013~hPa and a nominal height of the 150~g/cm$^2$ layer of 13.6~km, yielding
correction parameters of 
$ d(Rate) / dp=(-0.12 \pm 0.04)~ \% $/hPa and 
$ d(Rate) / dh=(-3.8 \pm 1.2)~ \% $/km.
This correction eliminates rate variations from the data-set, which are caused
by changing atmospheric conditions.  For a consistency check, a
rough estimate of the muon lifetime can be deduced from these values, assuming
that all muons are produced with the same energy at the same atmospheric depth.
The obtained lifetime of $ 2\pm0.5~\mu$s is consistent with the literature
value.

\section{Forbush Decreases} \label{events}

\begin{figure*}\centering
 \epsfig{width=\xx,file=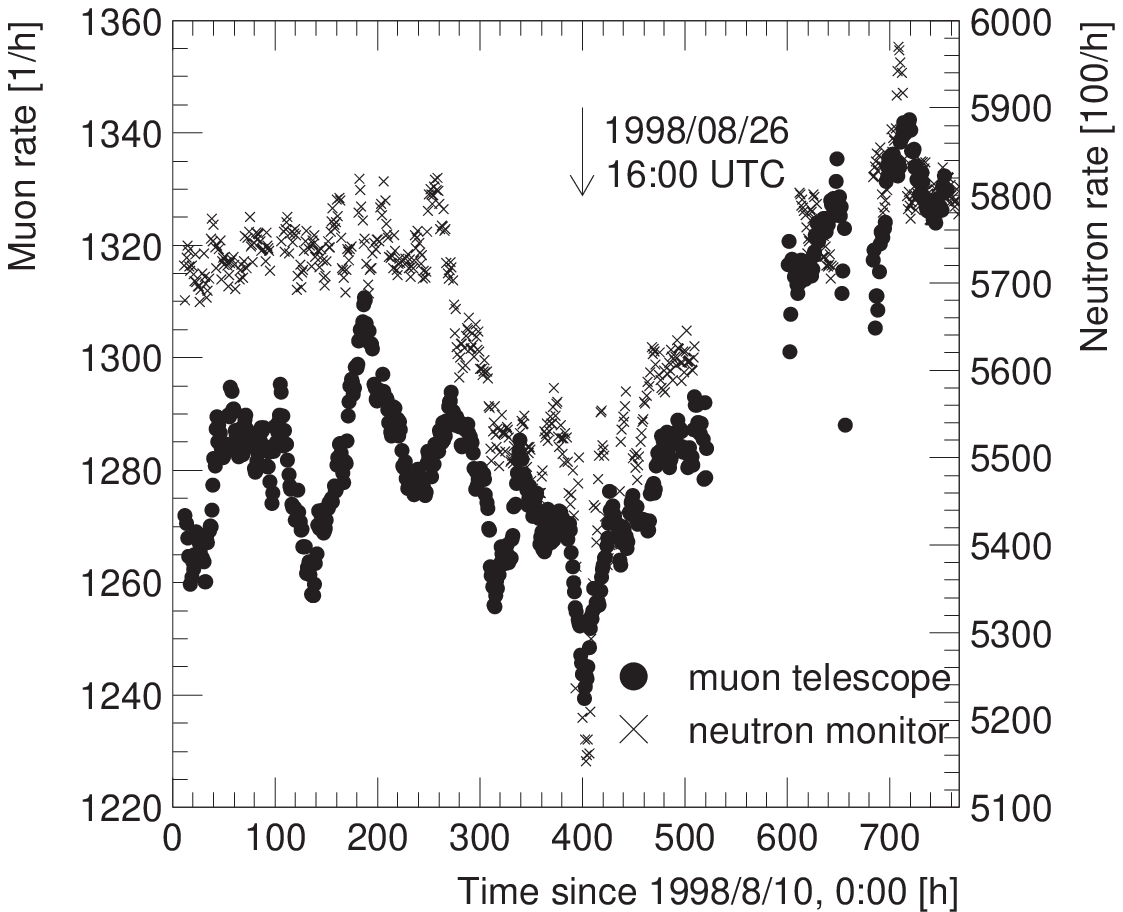}
 \epsfig{width=\xx,file=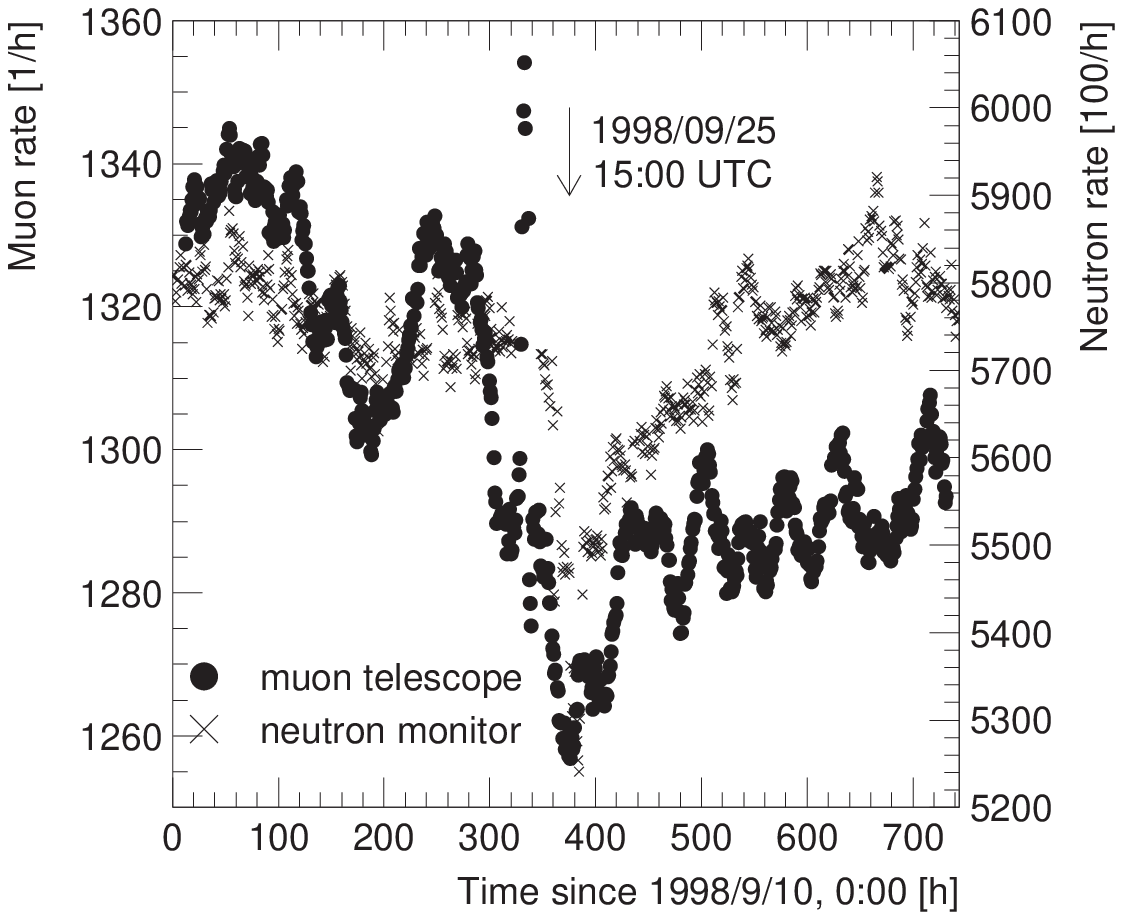}
 \epsfig{width=\xx,file=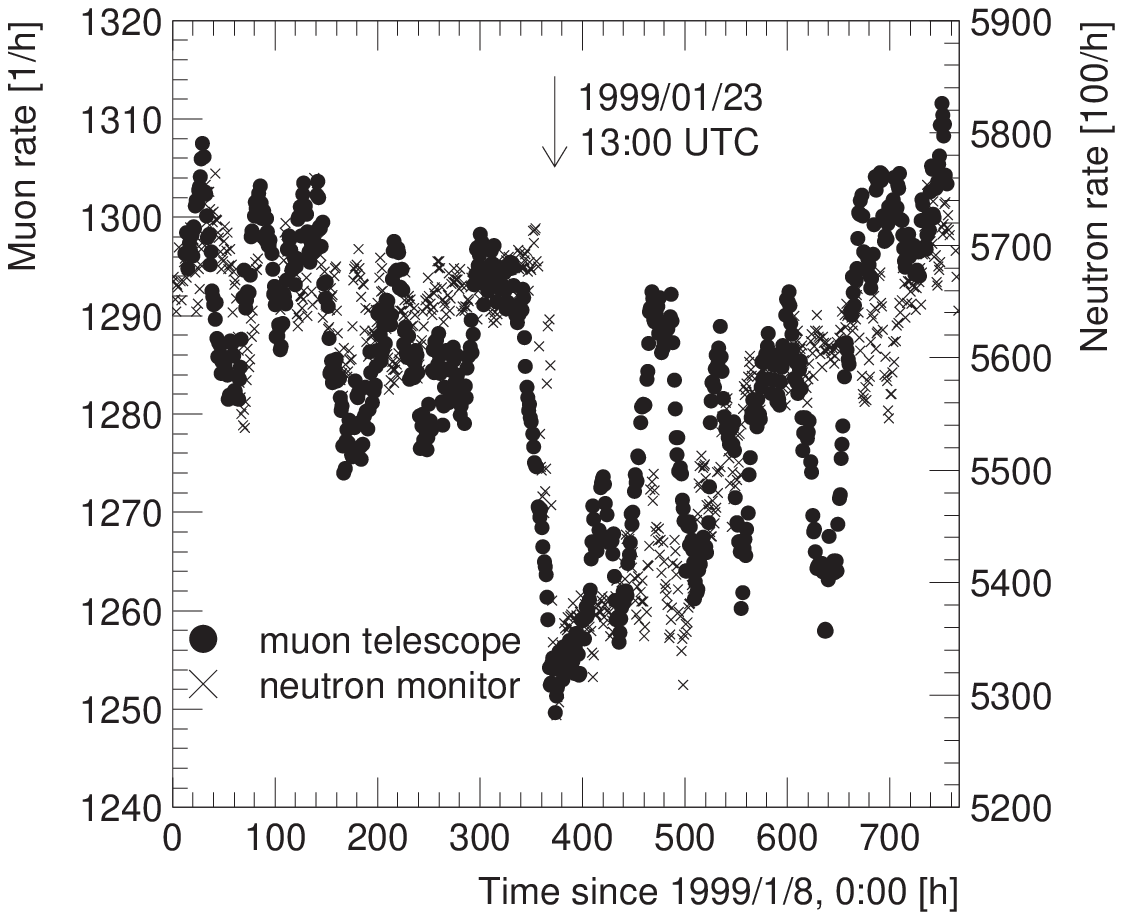}
 \epsfig{width=\xx,file=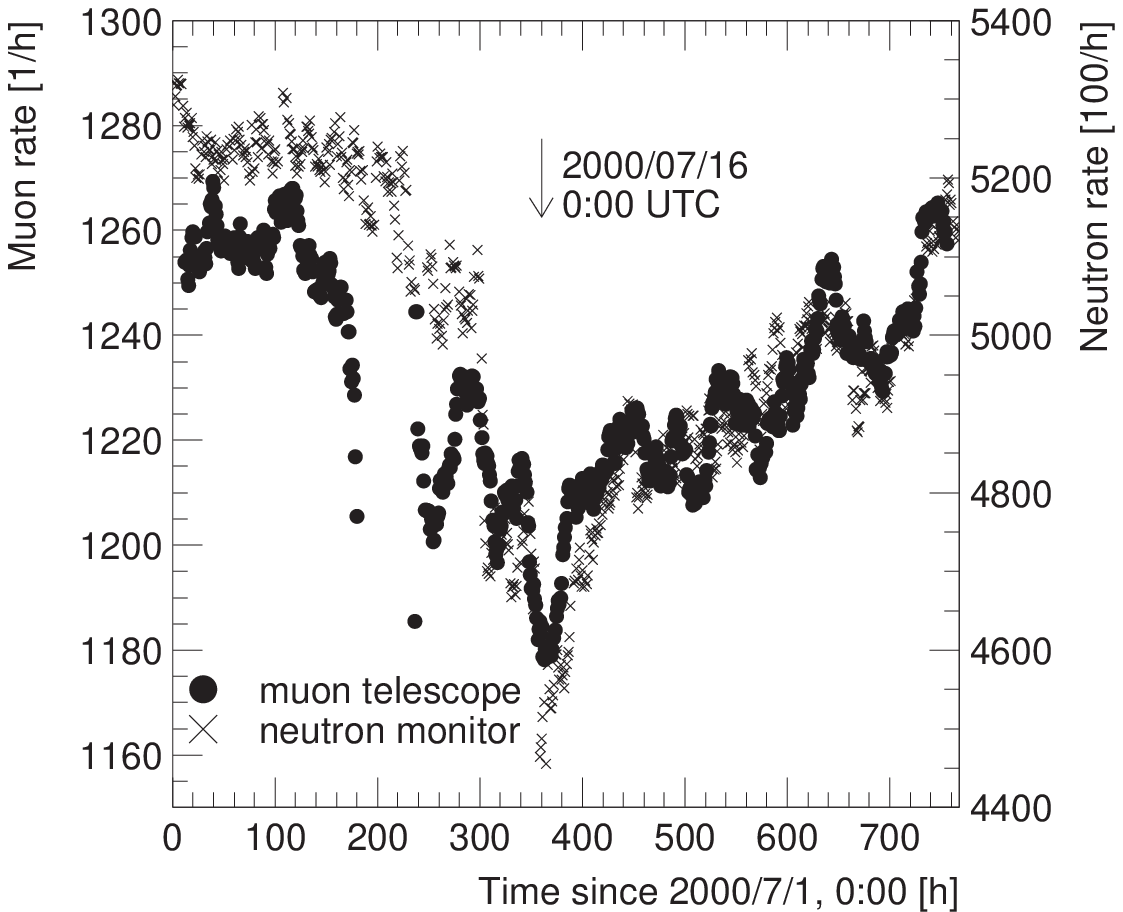}
 \epsfig{width=\xx,file=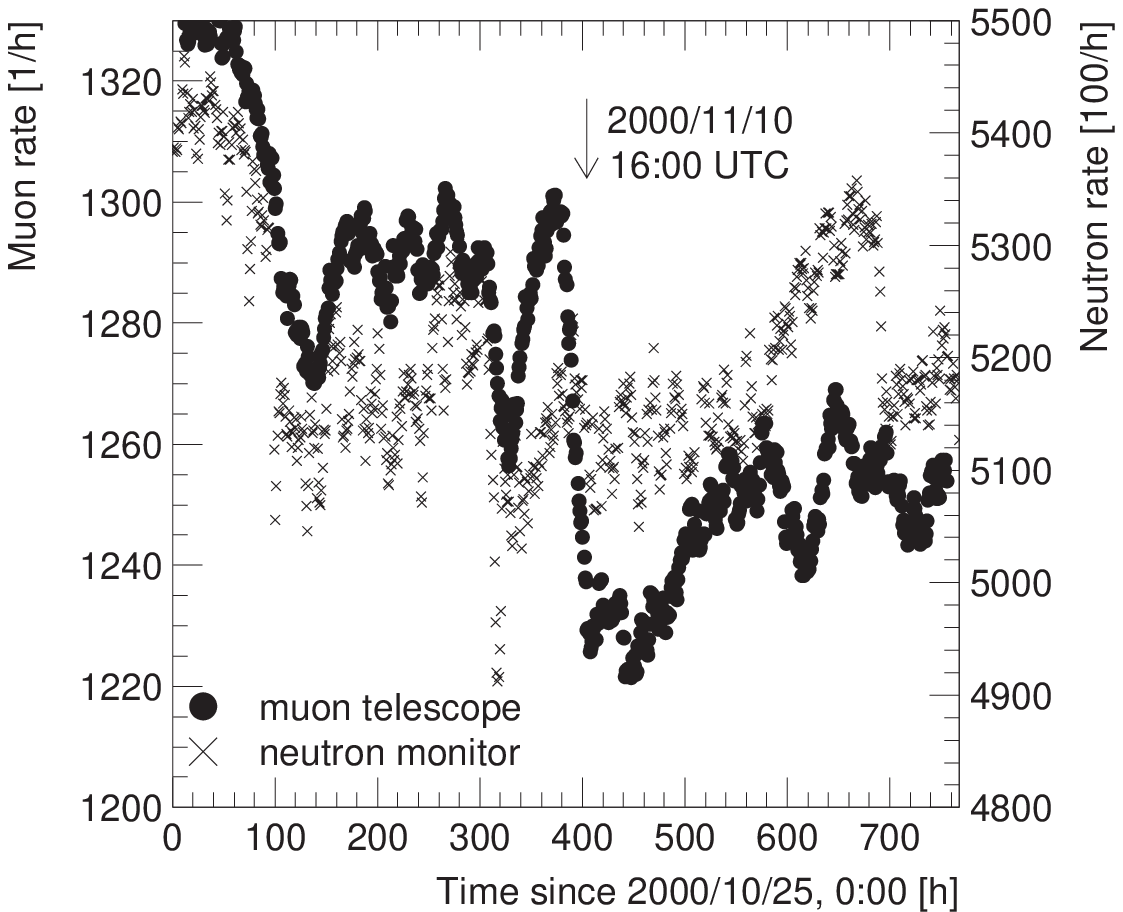}
 \epsfig{width=\xx,file=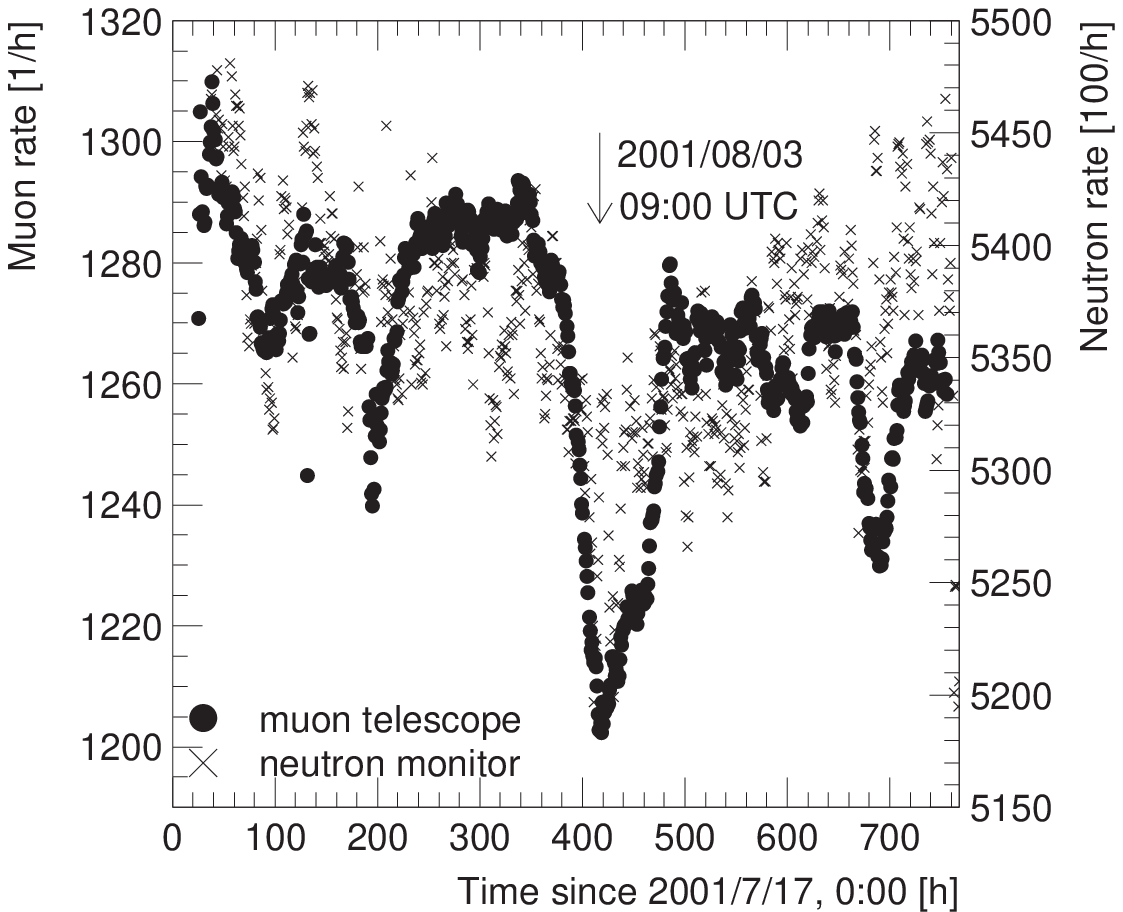}
 \caption{Count rates of the Karlsruhe Muon Telescope and the Jungfraujoch
          neutron monitor for several Forbush decreases. The corresponding
          dates and times are indicated in the figure.
	  Jungfraujoch data scaled by a factor 100, muon counting rate
          smoothed over a period of 24 hours. \label{events1}}
\end{figure*}

\begin{figure*}\centering
 \epsfig{width=\xx,file=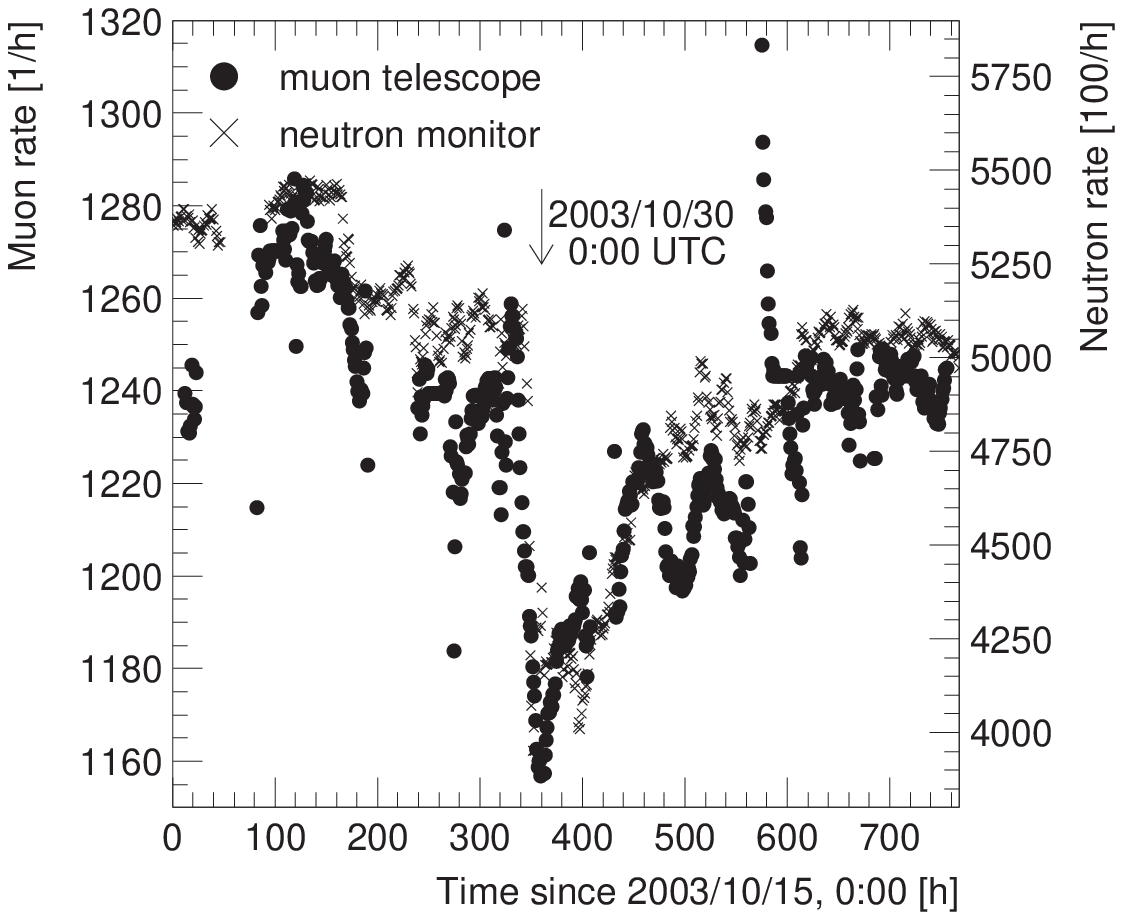}
 \epsfig{width=\xx,file=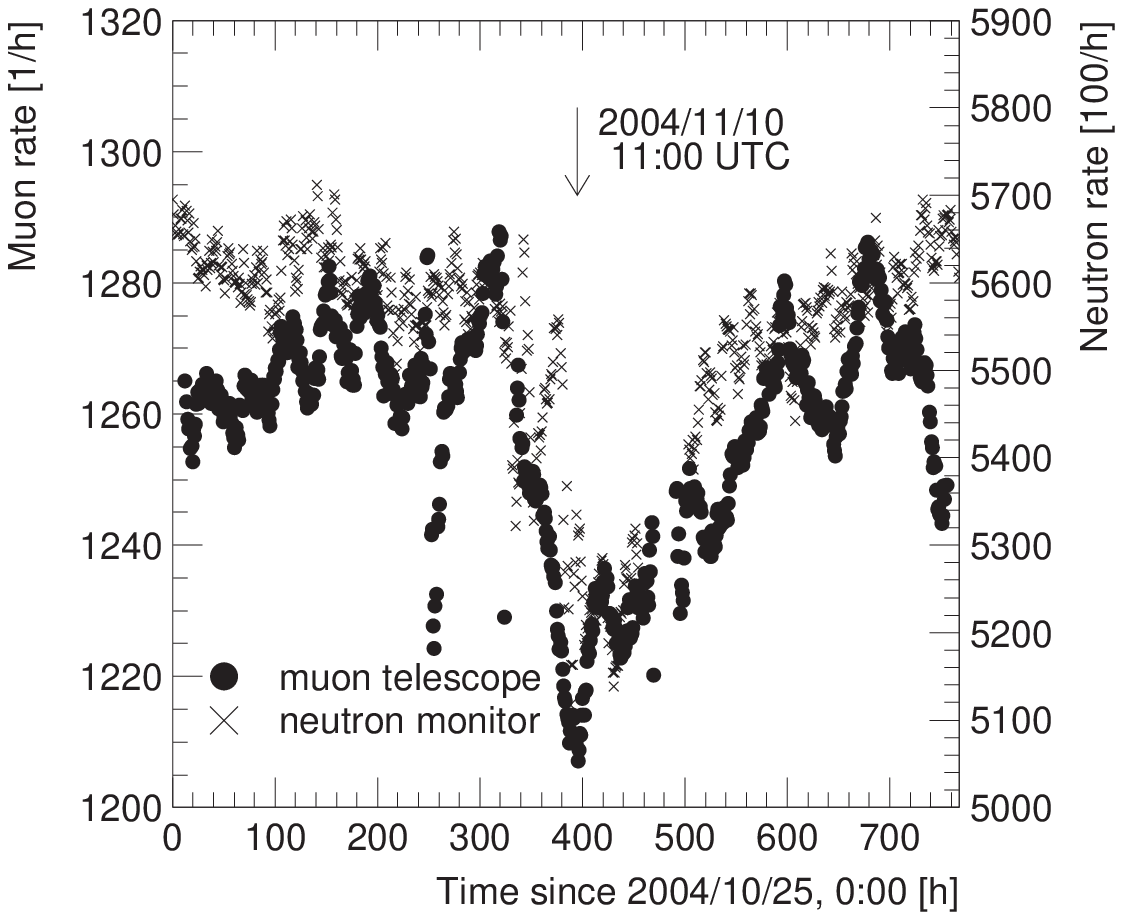}
 \epsfig{width=\xx,file=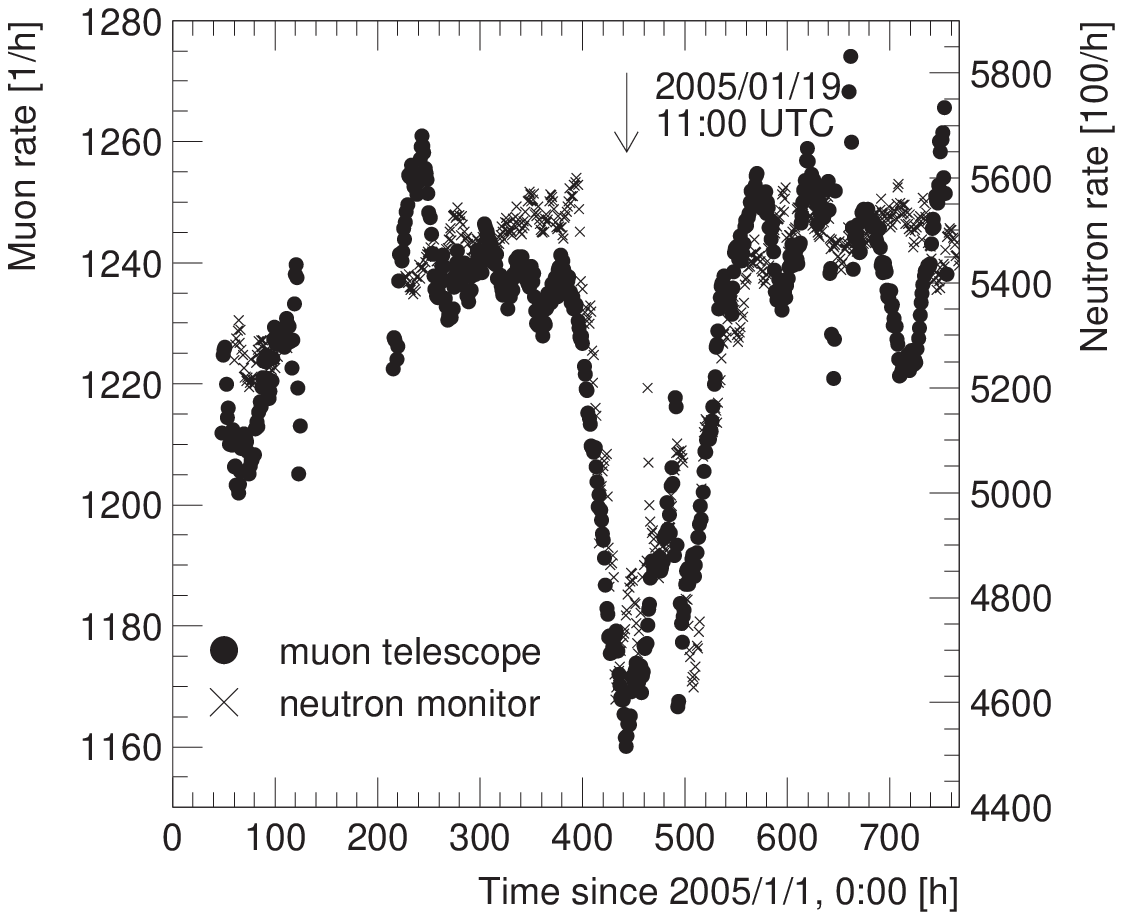}
 \epsfig{width=\xx,file=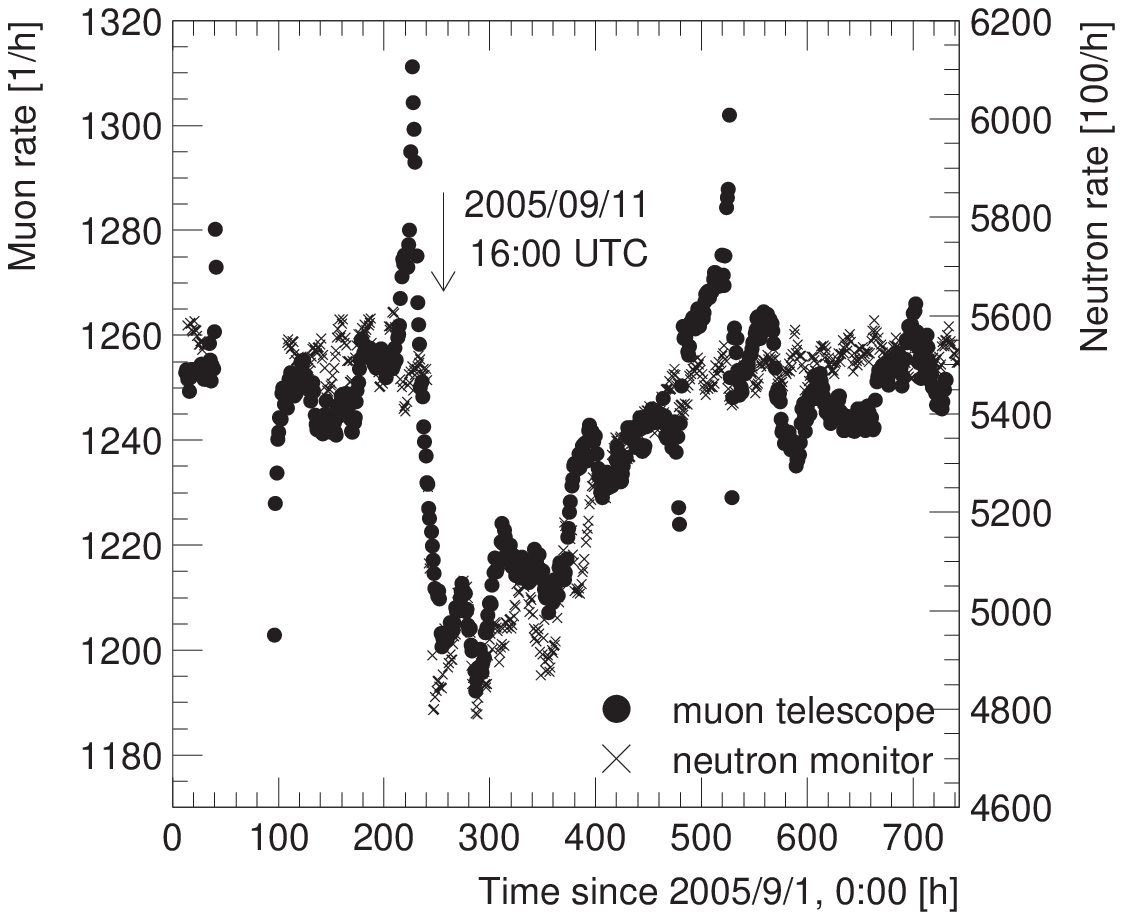}
 \caption{Count rates of the Karlsruhe Muon Telescope and the Jungfraujoch
          neutron monitor for several Forbush decreases. The corresponding
          dates and times are indicated in the figure.
	  Jungfraujoch data scaled by a factor 100, muon counting rate
          smoothed over a period of 24 hours. \label{events2}}
\end{figure*}

\begin{table*}\centering
 \caption{Very significant Forbush decreases detected since 1998. 
	  A sequential number and the dates are listed.  Significances are
	  pre-trials, the amplitudes of the Karlsruhe Muon Telescope refer to
	  the hourly data (not smoothed).  The fifth column gives the
	  amplitudes detected by the Jungfraujoch neutron monitor.  The last
	  column gives an estimate for the energy dependence of the detected
	  amplitudes, expressed in amplitude change per decade in energy.
 \label{tabevt}} 
 \begin{tabular}{rccccc}
 \hline
 \# &date &  significance($\mu$) & amplitude($\mu$) &  amplitude(n) & 
                                                             amplitude change\\
 & &   & [\%]  &  [\%] & [\% / energy decade]\\
 \hline
  1&1998/08/26 & 6.1$\sigma$ & 10.1 & 10.3 & -0.4\\
  2&1998/09/25 & 6.2$\sigma$ & 11.7 & 8.7 & 5.7\\
  3&1999/01/23 & 4.7$\sigma$ & 8.1  & 6.7 & 2.6\\
  4&2000/07/15 & 8.3$\sigma$ & 12.3 & 14.9 & -5.0\\
  5&2000/11/10 & 6.1$\sigma$ & 6.8  & 2.1 & 9.0\\
  6&2001/08/03 & 8.7$\sigma$ & 9.4  & 3.4 & 11.7\\
  7&2003/10/30 & 8.4$\sigma$ & 11.3 & 23.2 & -22.8\\
  8&2004/11/10 & 5.0$\sigma$ & 10.5 & 8.9 & 3.1 \\
  9&2005/01/19 & 10.6$\sigma$ & 13.2 & 16.8 & -6.9\\
 10&2005/09/11 & 5.5$\sigma$ &  8.4 & 13.5 & -9.8\\
 \hline
 \end{tabular}
\end{table*}

The muon data were searched for days where the average rate was significantly
lower than that of a background region. The background level was
determined from hourly count rates within two times two weeks (14~d before the
test region and 14~d afterwards), separated by three days from the tested day.
The significances for each day were computed according to \citet{LiMa}. Trial
factors were not taken into account.

The Karlsruhe Muon Telescope has detected several significant structures. The
strongest Forbush decreases in the years from 1998 to 2006 are compiled in
\tref{tabevt}. Shown are a sequential number, the date, the significance and
the amplitude $A$ of the minimum rate ($r_{FD}$) relative to the average rate 
before the decrease ($r_b$) computed as
\begin{equation}
A =(r_{b} - r_{FD}) / r_b.
\end{equation}
The amplitudes $A_\mu$ and $A_n$ have been calculated according to (1)
for the muon telescope (based on hourly
rates) 
and the Jungfraujoch 18-IGY neutron 
monitor (46.55$^\circ$N / 7.98$^\circ$E, 3570~m asl), respectively.
The latter has an effective vertical cutoff rigidity of 4.49~GV \citep{Bern}.
It was chosen for this comparison because of its geographic proximity to
Karlsruhe.  The detected events compared to the neutron monitor counting
rate are depicted in \ffref{events1} and \ref{events2}. To display their
development, the muon telescope rates are smoothed by a running mean over
24~hours.  Attention should be paid to the different scales for the muon rate
(left-hand scale) and for the neutron monitor rate (right-hand scale). 
The apparently significant excesses on 
1998/09/24, 2003/11/08, 2005/09/09-10 and 2005/09/22 are artefacts of 
the smoothing and caused by individual high data-points at the boundaries of 
detector down-time.
It is worth to point out that the rate development observed at 4.5~GV
(Jungfraujoch) and for the muon telescope (15~GeV) are quite similar, despite
of their different energy thresholds.
This illustrates that Forbush decreases are clearly detectable with a muon
detector with 15~GeV peak energy.  Forbush decreases were detected already with
the GRAND muon detector \citep{poirier11} at 10~GeV peak energy.  With the
Karlsruhe Muon Telescope we push the detection towards higher energies.

Many structures in these Forbush decreases are visible at both energies.  A
closer look reveals that for events 7, 8, 9, and 10 (close to the solar
minimum) the rates of both detectors follow each other extremely closely.
On the other hand, for events 1, 2, 4, and 6 there are systematic differences
between the two energies in the behavior before or after the Forbush decrease.
For the Forbush decrease in the year of the solar maximum (\# 5) the
strongest differences between the two rates are observed.
It appears as during solar maximum there are significant differences between
the fluxes observed at 4.5~GV and 15~GeV, while the fluxes are
correlated well during periods of low solar activity.

\begin{figure}\centering
 \epsfig{width=.8\columnwidth,file=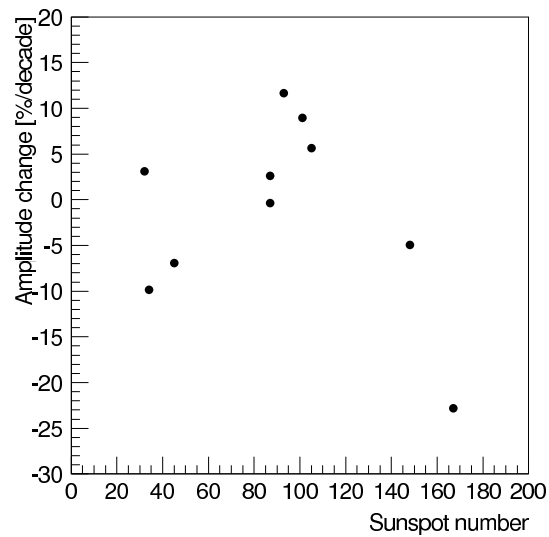}
 \caption{Energy dependence of the amplitude of the Forbush decreases
	  registered by the Karlsruhe Muon Telescope and the Jungfraujoch
	  neutron monitor (expressed as change in amplitude per energy decade)
	  as function of the sunspot number.}
 \label{assn}
\end{figure}

To study the energy dependence of the amplitudes of a Forbush decrease, the 
spectral index $\gamma$, i.e.
the change of amplitude per decade in energy has been calculated according to
\begin{equation}
\gamma = ({A_{\mu} - A_{n}})~/~{ ( log(E_{M}^{\mu}) - log(E_{M}^{n}))},
\end{equation}
$E_{M}^{\mu}\sim$ 15~GeV and $E_{M}^{n}\sim$ 4.5~GeV being the most propable 
primary energies for the muon telescope and the neutron monitor, respectively.
$\gamma$ is listed
in the last column of Table\,\ref{tabevt}. To investigate a possible dependence
on the solar activity, the amplitude change per energy decade is depicted as
function of the international sunspot number (taken from \citet{SIDC}) in
\fref{assn}. No clear correlation between the two quantities can be inferred
from the figure. Thus, earlier claims by \citet{ifedili} cannot be confirmed.
A study of the energy dependence of the recovery time of Forbush decreases
including data from Karlsruhe Muon Telescope is published elsewhere \citep{Uso08}.

\section{Ground Level Enhancements} \label{gle}

\begin{table}\centering
 \caption{Solar Ground Level Enhancements according to the Bartol group
          database \citep{Bartol}. The last column denotes the status
          of the Karlsruhe Muon Telescope:
          a: active, 
          i: inactive,
          +: GLE detected.
 \label{gletab}} 
 \begin{tabular}{ccl}
 \hline
 GLE number & event date   & status  \\
 \hline
     55 &   1997/11/06   &   i  \\
     56 &   1998/05/02   &   i  \\
     57 &   1998/05/06   &   a (+) \\
     58 &   1998/08/24   &   a  \\
     59 &   2000/07/14   &   a + \\
     60 &   2001/04/15   &   a + \\
     61 &   2001/04/18   &   a  \\
     62 &   2001/11/04   &   i  \\
     63 &   2001/12/26   &   i  \\
     64 &   2002/08/24   &   i  \\ 
     65 &   2003/10/28   &   i  \\ 
     66 &   2003/10/29   &   a \\
     67 &   2003/11/02   &   i \\
     68 &   2005/01/17   &   a  \\
     69 &   2005/01/20   &   a  \\
 \hline
 \end{tabular}
\end{table}

\begin{figure*}\centering
 \epsfig{width=\xx,file=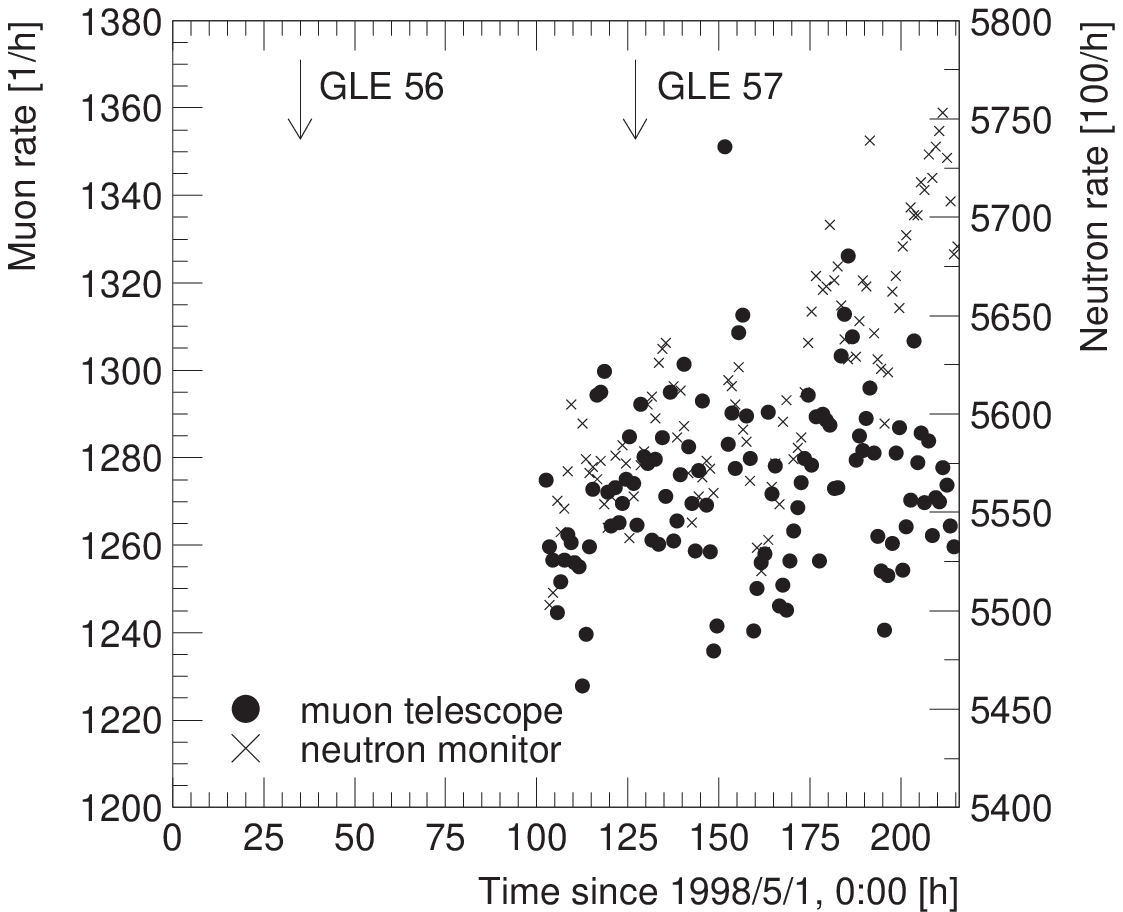}
 \epsfig{width=\xx,file=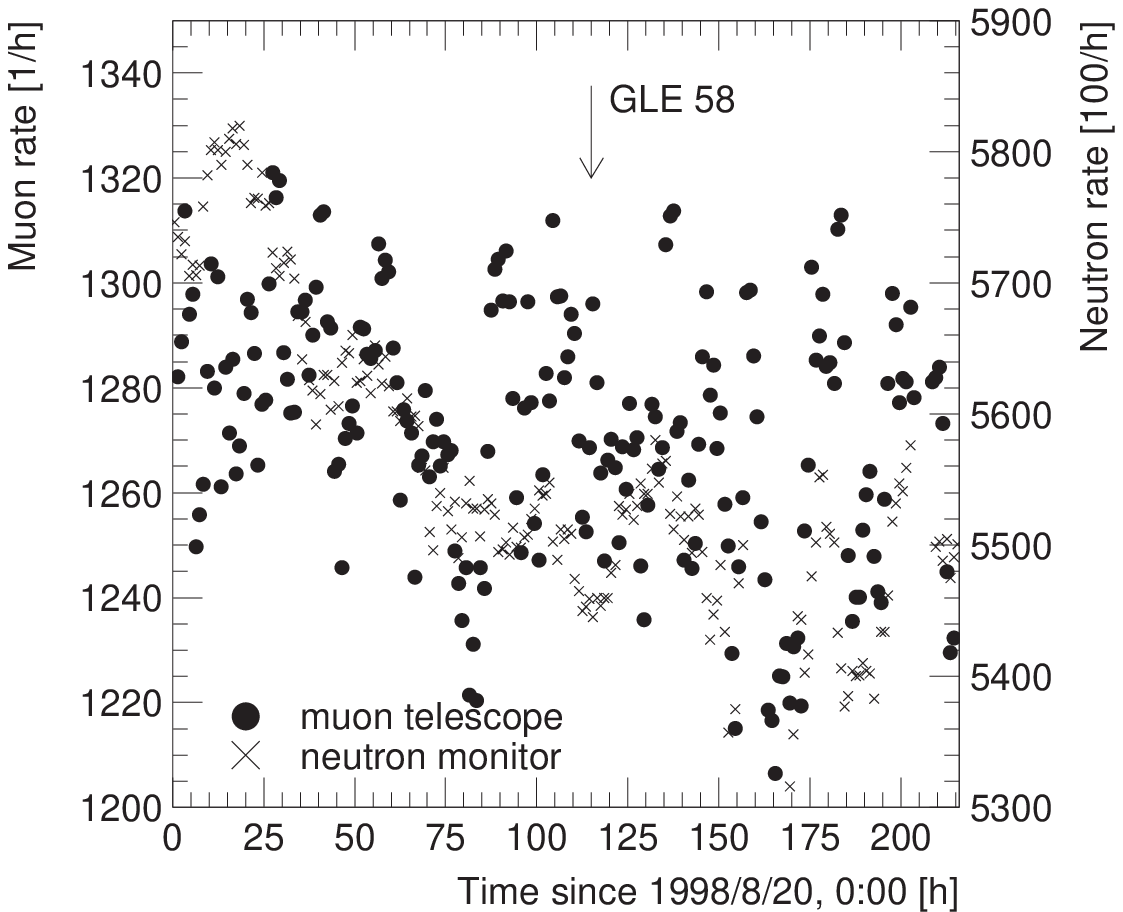}
 \epsfig{width=\xx,file=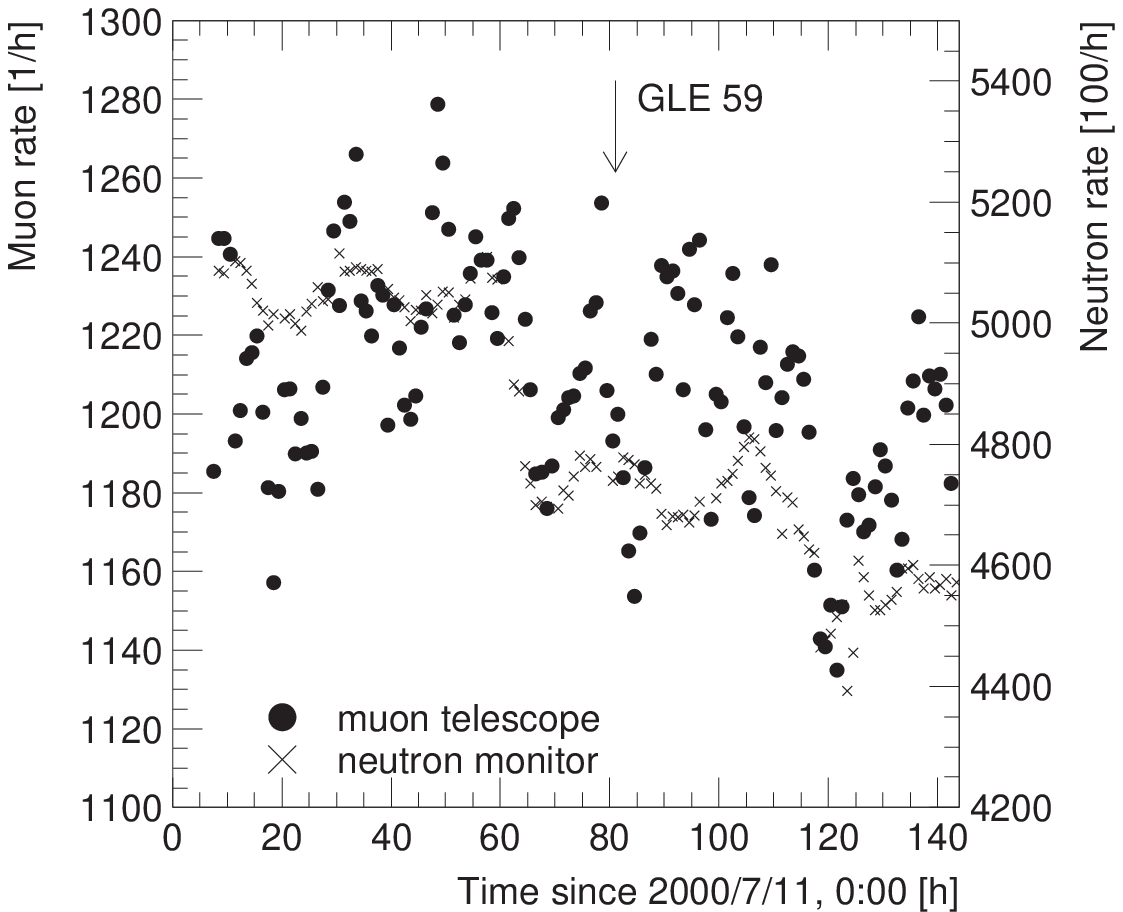}
 \epsfig{width=\xx,file=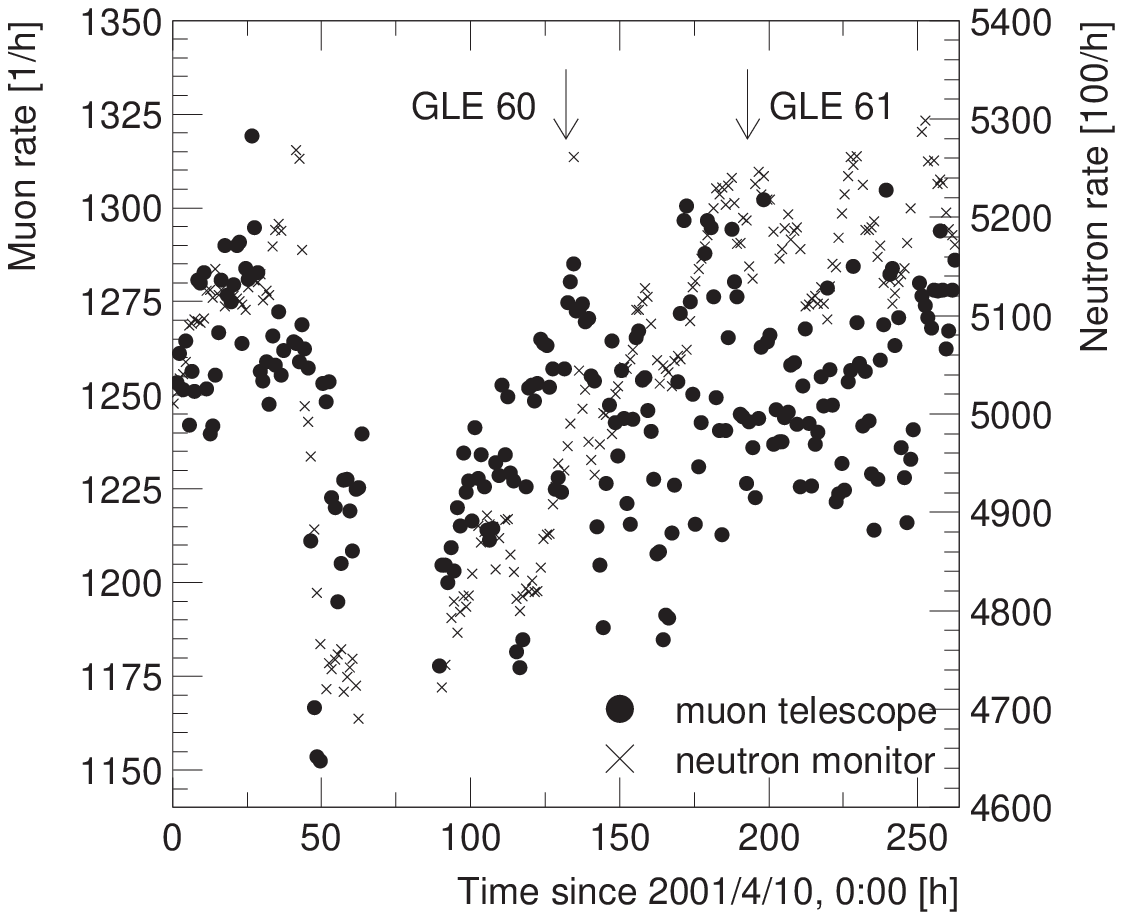}
 \epsfig{width=\xx,file=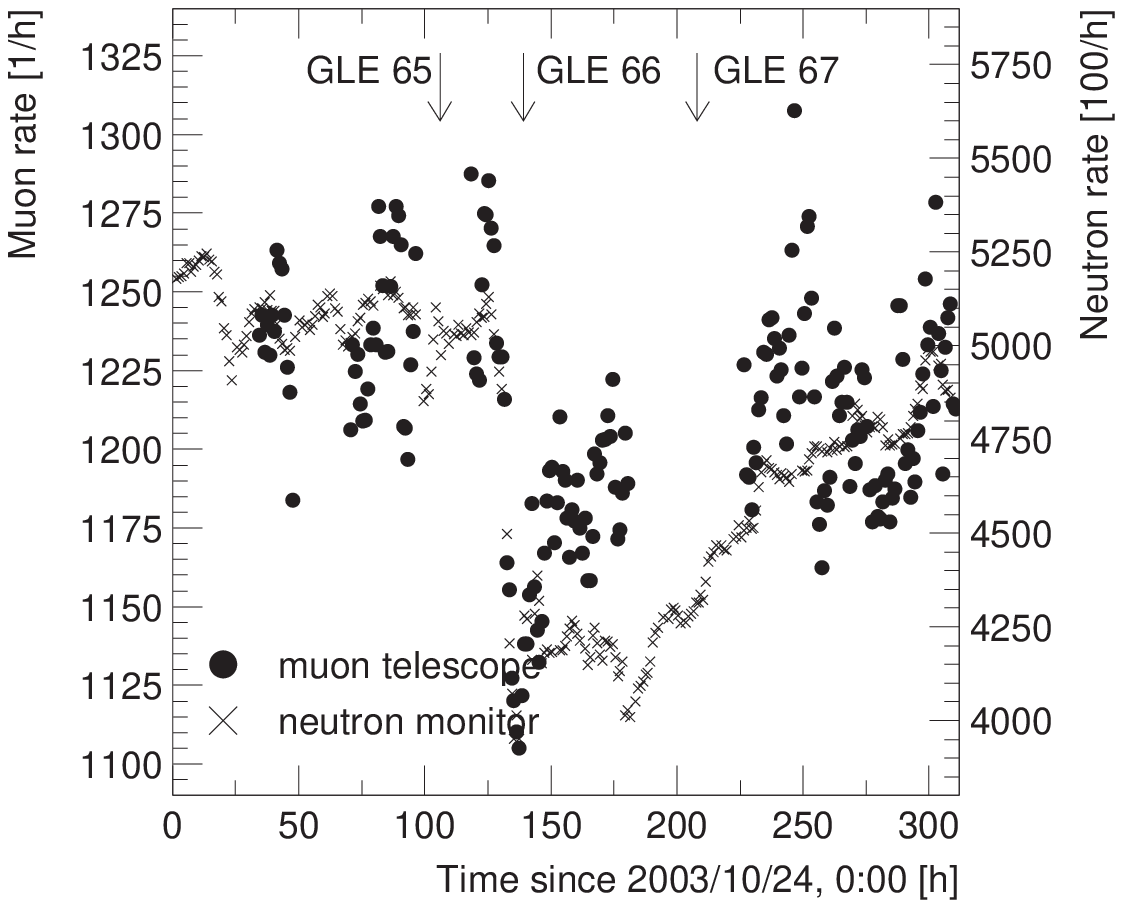}
 \caption{Hourly count rates registered by the Karlsruhe Muon Telescope and
	  the Jungfraujoch neutron monitor for several Ground Level
	  Enhancements, as marked in the figures, see also \tref{gletab}.
	  Jungfraujoch data are scaled by factor 100, muon counting rates
	  smoothed over a period of three hours.\newline
          \label{gle1}}
\end{figure*}

\begin{figure*}\centering
 \epsfig{width=\xx,file=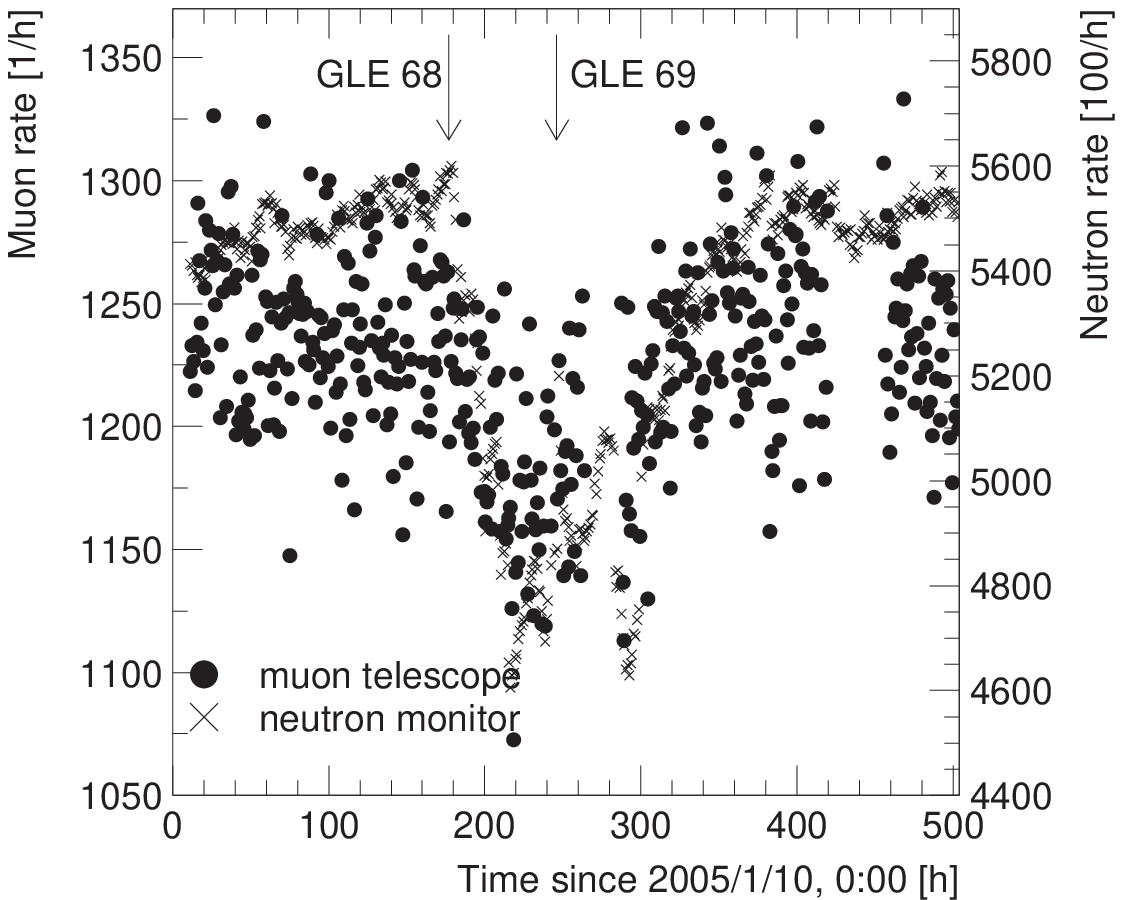}
 \epsfig{width=\xx,file=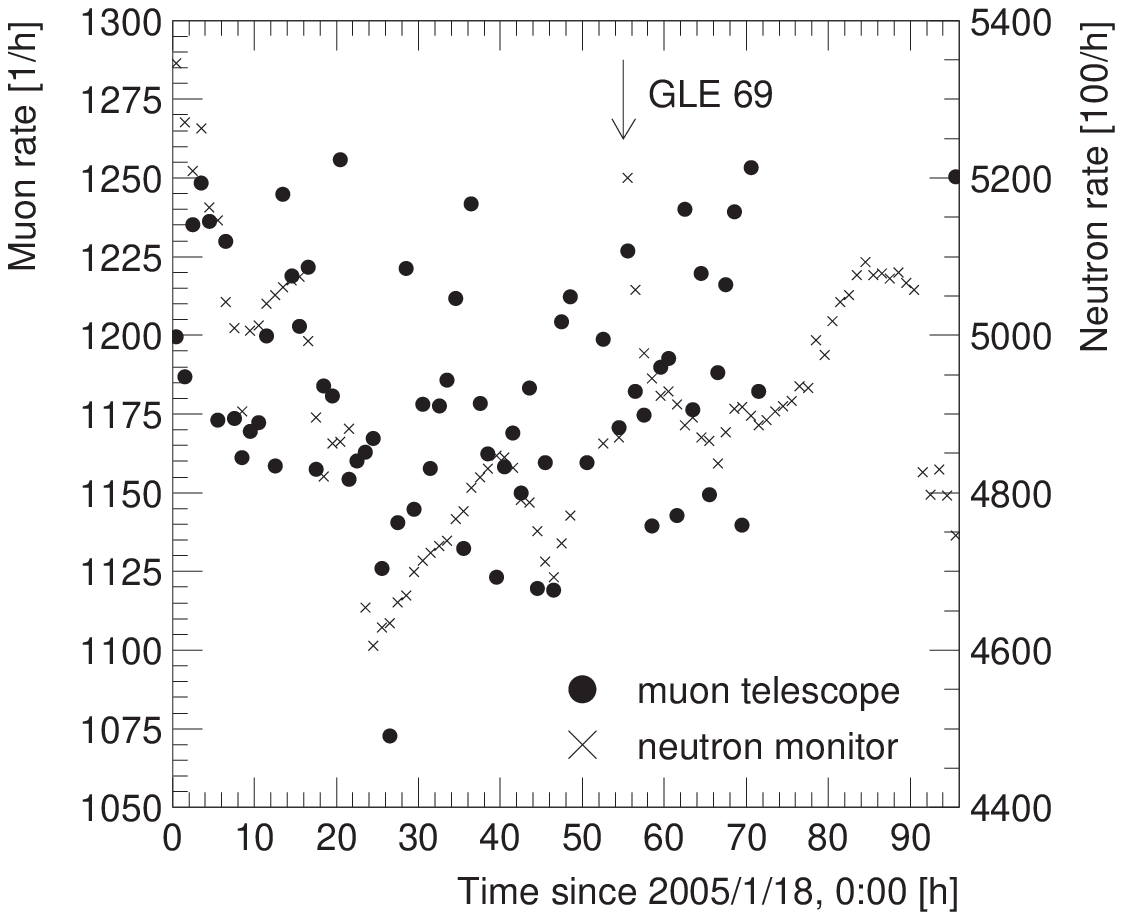}
 \caption{Count rates of the Karlsruhe Muon Telescope and the Jungfraujoch
	  neutron monitor for the Forbush Decrease on January 19$^{th}$, 2005.
	  Jungfraujoch data scaled by a factor 100, hourly (unsmoothed) muon
	  counting rate. The arrows indicate the positions of GLEs 68 and 69.
	  \emph{Right:} zoom into the region around GLE 69.
          \label{gle6869}}
\end{figure*}

Due to their relatively short duration, Ground Level Enhancements (GLEs) are
difficult to detect with the Karls\-ruhe Muon Telescope. Therefore, the data were
scanned for correlations with all events marked in the GLE database, as
provided by the Bartol group \citep{Bartol} and listed in \tref{gletab}. The
muon flux was recorded for events marked with an "a".  During events marked
with "i" the muon telescope was not active.  The hourly rates for GLEs 56 to 67
as registered by the Karlsruhe Muon Telescope and the Jungfraujoch neutron
monitor are depicted in \fref{gle1}. Jungfraujoch data are scaled by a factor
100 and the muon counting rates are smoothed over a period of three hours. 

For GLE 57, no significant excess was observed in the muon counting rate.
However, about seven hours before the Ground Level Enhancement a small peak is
visible in the registered muon flux. 
For GLE 58, no significant muon excess has been observed.

GLE 59, the "Bastille day event" on July 14, 2000  has been registered by many
detectors, including neutron monitors and space crafts 
\citep{bieber,vashenyukasr}. In
particular, the event could be measured for primary cosmic rays with GeV
energies \citep{wang}.  It has been detected by the GRAND muon detector (10 GeV
most probable energy) \citep{poirier14}, by the L3+C detector at CERN
($\approx40$~GeV primary energy) \citep{l3}, and also by the Karlsruhe Muon
Telescope. An excess in the muon counting rate can be recognized a few hours
before the event. The significance of this structure is under investigation. 
If real, it is a possible hint for energy dependent propagation effects or the 
strongly anisotropic nature of this event.

On Easter day 2001 (April 15) an event occured (GLE 60) which has been
observed and discussed by several groups
\citep{shea,dandrea,biebereaster,tylka,vashenyukasr}.
A muon count excess can be recognized at the time of GLE 60, while no signal is
observed from GLE 61. It should also be noted that the Jungfraujoch neutron
monitor detects GLE 60 with a large signal. On the other hand, the muon flux is
only slightly increased at the time of the event.

Some of the greatest bursts in the 23rd solar cycle occurred on 28/29 October
and 2 November 2003 (GLE $65-67$). They are extensively discussed in the
literature, \citep[e.g.]{watanabe,liu,hurford,eroshenko}.  
Unfortunately, the muon telescope was not active during GLEs
65 and 67. At the time of GLE 66, no significant signal is seen in the muon
count rate. However, about one day before GLEs 65 and 66 a peak can be
recognized in the registered muon flux. It is not clear if these increases are
statistically significant, since there are gaps in the observing time.  Thus,
it is not obvious if the detected rate variations are correlated with the
Ground Level Enhancements.

Unsmoothed hourly count rates of the Karlsruhe Muon Telescope compared to the
Jungfraujoch neutron monitor data during the Forbush Decrease in January 2005
are depicted in \fref{gle6869}.  For comparison the reader may refer to
\fref{events2} for the smoothed counting rate of the same event. In the
interval shown, two Ground Level Enhancements (GLE 68 and 69) have been
observed, the corresponding times are marked in the figure.  
GLE 69 occoured on January 20, 2005 and was the second largest GLE in fifty
years \citep{gle69-1,gle69-2,gle69-3,vashenyukasr}.
Measurements of the Aragats multidirectional muon monitor indicate that protons
were accelerated at the Sun up to energies of 20~GeV in this GLE
\citep{bostanjyan}. Protons accelerated during the main phase have a softer
energy spectrum than during the initial phase of the event.
It is assumed that protons were accelerated in a process or
processes directly related to a solar flare \citep{simnett}. 
The right-hand panel of \fref{gle6869} shows the region around GLE 69. No
indication for a significant increase in the muon rate associated with GLE 69
can be seen in the figure.
The Aragats data indicate that the time interval of solar proton flux with very
high energies was only very short, this could explain why nothing is seen in
the Karlsruhe Muon Telescope data. In addition, the solar cosmic-ray flux
during the initial phase was very anisotropic, another potential reason
for the non-observation in the muon rate.

\section{Conclusions} 

The Karlsruhe Muon Telescope provides information about effects of solar
activity on the cosmic-ray flux observed at Earth since 1993. The recorded muon
flux corresponds to 15~GeV peak energy (40~GeV median energy) for primary
protons.

Several strong Forbush decreases, i.e.\ a rapid decrease in the observed
galactic cosmic-ray intensity, could be measured with the muon telescope,
indicating that these effects can be seen at energies exceeding the typical
energies of neutron monitors. Comparing the observed amplitudes to the
Jungfraujoch neutron monitor data, the spectral index of the events has been
estimated. No dependence of the spectral index on the sunspot number has been
found. However, there are significant differences in the timely development of
the rates observed at 4.5~GV and 15~GeV for different states of solar activity.
For Forbush decreases during solar maximum, the rates of the muon telescope and
the neutron monitor behave quite differently, while they are well correlated for
periods of low solar activity.

It has been investigated whether Ground Level Enhancements, which are connected
to large solar flares, observed between 1997 and 2005 can be detected in the
registered muon flux. For the strong Ground Level Enhancements 59 and 60 a
clear signal can be seen in the muon count rate at the times of the events.
This provides direct evidence for particles being accelerated to energies as
high as 15~GeV during solar flares. Indirect evidence has been obtained
previously by observations of lines in the gamma ray spectrum measured during
solar flares \citep{rieger,fletcher}.  On the other hand, no signal has been
detected for the GLEs 58, 61, 66, 68, and 69.  If the underlying physics
processes of all Ground Level Enhancements are the same, this means that the
energy spectra of GLEs 59 and 60 differ from the spectra of the other GLEs.
Another possibility is that the angular distribution of the emitted particles is
different for different GLEs, i.e. in cases with highly anisotropic emission no
signal was detected in the muon counting rate.

\section*{Acknowledgments}

We are grateful to Mrs. Heike Bolz for her enthusiastic efforts in continuously
operating the Karlsruhe Muon Telescope and to J\"urgen Wochele for his help
during the construction of the detector.
We thank the team operating the Jungfraujoch neutron monitor for making their
data publicly available.
We acknowledge the help of the Deutscher Wetter-Dienst (DWD) and the Institut
f\"ur Meteorologie und Klimaforschung of Forschungszentrum Karlsruhe providing
atmospheric data.
We thank Jan Kuijpers for critically reading the manuscript and the anonymous
referees for useful advice.

\end{document}